# Quantum Cavitation[†]

Paola Zizzi*, Eliano Pessa* and Fabio Cardone**

\* Biokavitus Research Laboratory, Concesio (BS), Italy and University of Pavia, Italy. e-mails: paola.zizzi@unipv.it; eliano.pessa@unipv.it
\*\* Department of Physics "E. Amaldi", University of Roma 3, Italy; CNR-ISMN, Roma, Italy and Biokavitus Research Laboratory, Concesio (BS), Italy.

## Abstract

We consider the theoretical setting of a superfluid like $^3$He in a rotating container, which is set between the two layers of a type-II superconductor. We describe the superfluid vortices as a 2-dimensional Ising-like model on a triangular lattice in presence of local magnetic fields. The interaction term of the superfluid vortices with the Abrikosov vortices of the superconductor appears then as a symmetry breaking term in the free energy. Such a term gives a higher probability of quantum tunnelling across the potential barrier for bubbles nucleation, thus favouring quantum cavitation.

**Keywords**: Cavitation, superfluids, superconductors, quantum vortices, Ising model.

† Work partially supported by Biokavitus S.r.l. research grant.

## 1. Introduction

Cavitation is a complex physical phenomenon, typically involving the nucleation and growth of gas bubbles within a liquid. In presence of a positive pressure the bubbles undergo a very fast implosion, giving rise to strong energy concentrations, very high pressures and temperatures. The latter circumstances, which produce a number of harmful as well as advantageous effects, stimulated a deep study of cavitation, done by resorting to the theoretical apparatus of classical physics; see [1,2,3,4,5,6,7]. In recent times, however, both the evidence coming from experimental investigations about superfluid $^4$He and $^3$He and theoretical arguments showed the existence of phenomena of quantum cavitation [8, 9, 10, 11, 12, 13, 14, 15]. The latter consists in a strong increase of bubble nucleation rate induced by a quantum tunnelling across the nucleation barrier.

While the detailed theoretical description of this phenomenon heavily depends on the adopted choices both of the interatomic interaction potential and the form of the equation of state for the substance under consideration, it is very easy to understand why quantum effects should become important, with respect to thermal ones, at very low temperatures such as the ones associated with Helium superfluid phase. Indeed it is immediate to show that, for masses of the order of the nuclear mass of $^4$He or $^3$He and a temperature of the order of 0.01 °K, the thermal De Broglie wavelength has a value of the order of $10^{-10}$ m, just of the same order of parameter which, in a Lennard-Jones interatomic potential, characterizes the minimal interatomic distance. We remark, however, that the above argument could hold even in other situations, related to cavitation but not necessarily associated with low temperatures. For instance, in the final stage of bubble collapse in a cavitation it has been reported [16] [17] that temperatures of $10^5$ °K (and higher) could be reached.

However, despite the technological difficulties, the effects occurring at very low temperatures are still easier to investigate than the ones associated with very high temperatures. Thus, in this paper we will be concerned about quantum cavitation in a superfluid (typically $^3$He) at temperatures very close to the absolute zero.

We will consider a superfluid like $^3$He contained in a vessel which is rotating at a high angular velocity. In this state of the superfluid there is the formation of quantum vortices whose pattern is a triangular lattice. The presence of vortices gives a further contribution of negative pressure, favouring the formation of bubbles (see, for instance, [18] [19]).

Also, we will take into account the influence of Abrikosov vortices occurring in a type-II superconductors (which also form a triangular lattice) on superfluid vortices.

We remark that this interaction between the two lattices of vortices appears as natural (owing to their similarity and the common theoretical background underlying the description of superfluidity and superconductivity). Also, we expect that this interaction will increase the cavitation rate.

Insofar, in the literature, this kind of interaction between a superconductor and a superfluid has been taken in consideration only within the context of neutron stars (see, for instance [20]) or of high-density nuclear matter [21].

In Sect. 2, neglecting obvious experimental difficulties, we will give the theoretical setting of a possible experiment, where the lattice of the Abrikosov vortices is superposed to the lattice of superfluid vortices. This can be modelled as a simple Ising-like model in 2D on a triangular lattice. Sect. 3 is devoted to the conclusions.

## 2. The model

We give the theoretical setting of an experiment which might improve quantum cavitation.

Let us consider a superfluid (for example $^3$He) put in a rotating container. To be a superfluid, helium is cooled at a temperature below the $\lambda$-temperature: $T < T_\lambda$. For low angular velocities $\Omega$ of the container, the superfluid is stationary: $v_s = 0$, where $v_s$ denotes the superfluid velocity. This is called the Landau state. Therefore, the superfluid in the Landau state is irrotational (**curl** $v_s = 0$).

But when Ω increases, the Landau state becomes unstable and decays toward a state having vortices.

In this state, the bulk of the superfluid remains irrotational. There are however thin regions of the superfluid, which circulate around vortices lines, in the same rotational sense of the container. The superfluid vortices form a triangular lattice.

Now, let us imagine to put the container between the two layers of a type-II superconductor, with an external magnetic field $H$ greater than the critical value: $H > H_c$. As it is well known, in this case type-II superconductors have Abrikosov vortices, which also form a triangular lattice. We arrange things in such a way that the two lattices superpose one to another, in order to look for the interaction between the two kinds of vortices (See Fig.1).

**Fig.1**
**A superfluid in a rotating container**
**sandwiched in a Type-II superconductor.**
**The yellow straight lines are superfluid**
**vortex lines. The green curled lines are**
**Abrikosov vortex lines.**

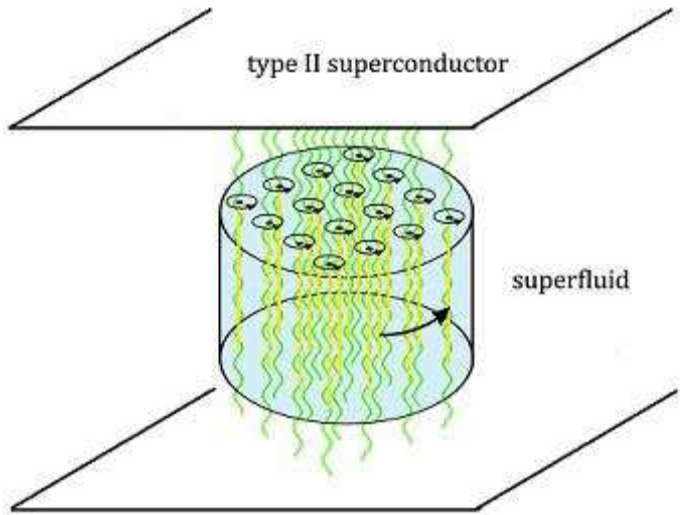

Let us consider a particle (an atom) of the superfluid with mass $m_s$, which is circulating around a closed loop surrounding a vortex line at a certain distance $\vec{r}$ from the core of the vortex with tangential velocity $\vec{v}$.

In the case of a superfluid, the circulation of the velocity field $\vec{v}$ along an oriented closed path C enclosing a vortex line is quantized:

$$\Phi = \oint_C \vec{v} \cdot d\vec{l} = \frac{2\pi\hbar}{m_s} n \tag{1}$$

where $d\vec{l}$ is the infinitesimal element of C, and $\hbar$ is the reduced Planck constant: $\hbar = \dfrac{h}{2\pi}$.

It follows that each superfluid atom near the vortex has one unit of quantized angular momentum, and a velocity:

$$v(r) = \frac{\hbar}{m_s r} \tag{2}$$

The vorticity of the particle is:

$$w(r) = \frac{2\hbar}{m_s r^2}. \tag{3}$$

In a sense, vorticity is a measure of the local spin of part of the superfluid.

Our purpose is to find an interaction between the superfluid vortices described above and the Abrikosov vortices of a type-II superconductor, whose associated quantized flux is :

$$\Phi = \oint_S \vec{B} \cdot \hat{n} \, dS = \frac{2\pi\hbar}{e_c} n \tag{4}$$

where $e_c$ is the charge of the Cooper pairs, and $\hat{n}$ is the unit vector normal to the surface.

The magnetic field distribution of a single vortex:

$$B(r) \approx \sqrt{\frac{\lambda}{r}} e^{-\frac{r}{\lambda}}$$

where r is the distance from the core and $\lambda$ is the London penetration depth, logarithmically diverges for $r \to 0$. However, in practice, for $r \leq \xi$, where $\xi$ is the coherence length, we have:

$$B(0) \approx \frac{\Phi}{2\pi\lambda^2} \ln k \tag{5}$$

where $k = \frac{\lambda}{\xi}$ is the Ginzburg-Landau parameter.

From (4) and (5) one gets:

$$B(0)_{ABRIK} \approx \frac{\hbar}{e_c \lambda^2} \ln k \tag{6}$$

Both kinds of vortices (1) and (4) form triangular lattices, which can be superposed one to another. The simplest theoretical model which can describe this interaction is an Ising-like model in 2D in presence of local magnetic fields. In this case, spins (defined from vorticity) are attached to the nodes of the lattice. Moreover, local, quantized magnetic fields (defined from Abrikosov vortices) interact with spins at each node. Within this description, it is evident that the sites of the lattice coincide with the cores of the vortices. The cores are quite unapproachable because in the vicinity of the vortex core, the velocity $\vec{v}$ is not defined. However, in presence of a magnetic field $\vec{B}$ of suitable intensity, it might be possible to get the superfluid particle at a distance of the order of the atomic scale ($r_0 \approx 10^{-10} m$) from the vortex core.

Then, let us introduce a magnetic field $\vec{B}$ having the same direction and verse of the vorticity $\vec{\omega}$ along the vortex line. If the superfluid particle has an electric charge $q_s$, it will experience a magnetic force $\vec{F} = q_s (\vec{v} \times \vec{B})$. The force causes the particle to be angularly displaced, i.e. spiral.

By equating the centripetal force and the magnetic force, both evaluated at the cut-off $r_0$, one gets:

$$B_{CYCL} = \frac{\hbar}{r_0^2 q_s} = 0.3 \times 10^5 \, Wb/m^2 \tag{7}$$

In correspondence to the cut-off, the cyclotron frequency is:

$$f_c(r_0) = 3 \times 10^8 \, Hz \tag{8}$$

At the cut-off, the particle is "almost" spinning around its axis, which "almost" coincides with the vortex line, and its angular momentum is not undefined, and it is still quantized.

Then, at the cut-off, we are authorized to build up an Ising-like model where we identify the spins $S_i$ at each site $i$, with vectors proportional to the vorticity $\vec{w}$ at each vortex core, that is, we put:

$$S_i = \eta w_i \qquad \text{with:} \quad \eta = \frac{r_0^2 m_s}{2}. \qquad (9)$$

We remind that the magnetic field B considered here should be a local field related to an Abrikosov vortex of a type-II superconductor. By comparing the intensity of the Abrikosov vortex magnetic field in (6) with that of the cyclotron magnetic field in (7) one gets:

$$\ln k = \gamma \qquad (10)$$

with:

$$\gamma = \frac{\delta}{\rho}, \quad \delta = e_c \lambda^2, \quad \rho = q_s r_0^2 \qquad (11)$$

Then, for our purpose, we need a type-II superconductor with a Ginzburg-Landau parameter of value $k = \exp(\gamma)$.

The above considerations allow us to write the free energy of an Ising-like model on a triangular lattice in terms of the $w_i$, in presence of localized magnetic fields $h_i$:

$$E = -J\eta^2 \sum_{ij}(w_{i,j}w_{i,j+1} + w_{i,j}w_{i+1,j} + w_{i,j}w_{i+1,j+1}) - \rho\eta \sum_i h_i w_i \qquad (12)$$

where $J$ stands for the common value of all the $J$s, and is taken positive (ferromagnetic interaction). The constants $\eta^2$ and $\rho\eta$ placed in front of the spin-spin and magnetic field-spin interaction terms respectively, are required by dimensional arguments.

The sum in the spin-spin interaction term links every site to its nearest neighbors. See Fig.2.

**Fig.2**
**The triangular lattice of vorticity**
**in a Ising-like model**

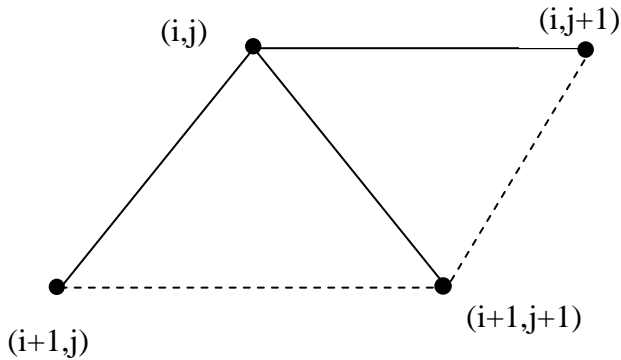

As in the original Ising model in 2D, the magnetic fields break the symmetry under the interchange of +1 (spins aligned) and -1 (spins anti-aligned).

For $J \approx 0$ (almost non interacting superfluid vortices), the first term in (12) can be neglected and the leading term is the interaction term between superfluid vortices and Abrikosov vortices:

$$\Delta E \approx -\rho\eta \sum_i h_i w_i \qquad (13)$$

Cavitation occurs in superfluids at negative pressures where the energy barrier preventing the nucleation of bubbles becomes smaller.

The negative pressure at which the compressibility of the liquid becomes infinite, the sound velocity becomes zero and the barrier to nucleation vanishes is called the spinodal limit.

At some distance from the spinodal, there is a chance for bubbles nucleation with the aid of thermal fluctuations. At low temperatures, however, thermal fluctuations become weaker, and the probability of nucleation remain small even very close to the spinodal. Below a critical (crossover) temperature T*, thermally activated nucleation becomes negligible compared to quantum cavitation (quantum tunnelling through the energy barrier). For thermally assisted quantum tunnelling see for example [22].

The cavitation rate in a superfluid (that is, the number of nucleated bubbles per unit time and volume) evaluated at the thermal-to-quantum crossover temperature T* is [9]:

$$J_{T^*} = J_{0T^*} e^{-\Delta\Omega_{max}/T^*} \tag{14}$$

Where $\Delta\Omega_{max}$ is the energy gap and $J_{0T^*}$ is a factor depending on the particular nucleation process.

In presence of interaction of superfluid vortices with Abrikosov vortices, the cavitation rate is then increased by a factor:

$$e^{-\Delta E} = e^{\rho\eta\sum_i h_i w_i} \tag{15}$$

Notice that the contribution to quantum cavitation given in (15) does not depend on temperature, but is purely quantum.

### 3. Conclusions

In this paper, we have considered a Gedankenexperiment, where a superfluid contained in a vessel rotating at high angular velocity, is sandwiched between the two layers of a type-II superconductor. In this experimental setting, we pretend to be able to arrange things in such a way that the two triangular lattices of Abrikosov vortices and superfluid vortices coincide. In this way, we can build a mathematical model which is an Ising-like model in 2D on a triangular lattice in presence of local magnetic fields. This leads to the deduction of formula (15) which gives a purely quantum contribution to quantum cavitation. By purely quantum we mean that this contribute to the cavitation rate does not depend on temperature, differently from the usual thermally assisted quantum cavitation. The fact that this term is temperature independent is quite important because it can be applied to several different domains. We wish to remark that this term indicates that a quantum transition is taking place. As it is well known, quantum transitions occur without divergence of fluctuations close to the critical point. Moreover, quantum phase transitions are associated to the so called quantum order, which is purely topological [23].

The formula expressing the topological contribution to quantum cavitation can be read as describing a new kind of quantum effect, taking place in some many-body systems endowed with topological singularities associated with quantized fluxes. This effect consists in the occurrence of a further negative pressure gradient whenever another quantized flux, related to an external field characterized by similar topological singularities, is coupled with the flux associated with the many-body system under consideration. This pressure gradient is, within some limits, independent from the system temperature and its effect is mediated by the spatio-temporal volume in which the coupling acts. In any case, it enhances in a very significant way the nucleation rate of bubbles within the system. In turn, this gives rise to a new form of quantum cavitation, which can produce energy release either through an implosion of these bubbles or through their explosion. In both cases we have the formation of a microplasma, whose origin in entirely due to a quantum effect, offering an an alternative with respect to other traditional ways of plasma formation. While this effect has been described within the specific context of superconductor-superfluid interaction, we hypothesize that it is endowed with a more general nature and can therefore take place even in other contexts. This hypothesis is supported by the fact that Ising-like models are conceptual structures having a wide reach, being used to describe very different systems, ranging from ferromagnets to biological neural networks, from crystal structures to cognitive systems.

The phenomena associated with the microplasma quoted above could be the cause of observed piezonuclear effects both in liquids and solids (see, for instance, [24, 25, 26, 27]). While the domain of nuclear phenomena in condensed matter is still associated with contradictory experimental evidence (see for a critical discussion [28]) quantum cavitation and its associated microplasma could offer the basis for a useful model to understand this very complex phenomenology.

**Acknowledgements**

We wish to thank Andrea Rampado, President of Biokavitus, for his interest in the present work, support and encouragement.